\documentclass[12pt,a4paper]{iopart}
\usepackage{ifpdf}
\usepackage[utf8]{inputenc}
\usepackage[english]{babel}
\usepackage{rotating}
\usepackage{graphicx}
\usepackage{color}
\usepackage{relsize}

\definecolor{reddeep}{rgb}{0.995,0.000,0.200}

\begin{document}
\title
{Phase Transition of the Uniaxial Disordered Ferroelectric Sr$_{0.61}$Ba$_{0.39}$Nb$_2$O$_6$.}
\author{
S.N. Gvasaliya$^1$, R.A. Cowley$^2$, L.I. Ivleva$^3$, S.G. Lushnikov$^4$, B. Roessli$^5$, and 
A. Zheludev$^1$
        }
\address{$^1$ Laboratory for Solid State Physics, ETH Z\"urich, CH-8093 Z\"urich, Switzerland\\
         $^2$ Clarendon Laboratory, Department of Physics, Oxford University, Parks Road, Oxford, OX1 3PU, UK\\
         $^3$ Prokhorov General Physics Institute, 119991, Moscow, Russia\\
         $^4$ Ioffe Physico-Technical Institute RAS, 194021, St. Petersburg, Russia\\
         $^5$ Laboratory for Neutron Scattering, Paul Scherrer Institut, Villigen PSI, CH-5232, Switzerland\\
}
\ead{sgvasali@phys.ethz.ch}
\begin{abstract}
We report a neutron scattering study of a ferroelectric phase transition in Sr$_{0.61}$Ba$_{0.39}$Nb$_2$O$_6$ (SBN-61). 
The ferroelectric polarization is along the crystallographic $c$-axis but the transverse 
acoustic branch propagating along the $<$1, 1, 0$>$ direction does not show 
any anomaly associated with the this transition. We find no evidence for a soft transverse optic phonon. 
We do, however, observe elastic diffuse scattering. The intensity of this scattering increases as the sample 
is cooled from a temperature well above the phase transition. 
The susceptibility associated with this diffuse scattering follows well the anomaly of the dielectric permittivity of SBN-61. 
Below T$_\mathrm{c}$ the shape of this scattering is consistent with the scattering expected from ferroelectric domain walls. 
Our results suggest that despite apparent chemical disorder SBN-61 behaves as a classic order-disorder 
uniaxial ferroelectric with critical fluctuations in the range $<10^{-11}$~s. 
\end{abstract}
\maketitle
\section{Introduction}

\noindent 

Relaxor ferroelectrics are disordered crystals with an anomaly in the dielectric permittivity $\varepsilon$ that is 
broad in temperature and frequency-dependent. Remarkably, this anomaly does not necessarily link to any macroscopic changes of 
symmetry~\cite{cross_1987}. 

Due to very high values of $\varepsilon '$, large piezoelectric constants and electro-optic coefficients, the 
relaxors have been a focus of intense experimental and theoretical 
studies~\cite{PhysRevLett_111_097601_levanyuk},~\cite{PhysRevLett_110_147602_rappe},\cite{2013arXiv1309.2816G},\cite{advphys_60_229}.  
It is commonly accepted that the unique properties of relaxors are related to the disordered chemical structure. 
However, a consistent model of the relaxor behavior has not so far been developed. In particular, it is unclear how large the chemical 
disorder has to be for a ferroelectric material to become a relaxor. We report below a neutron scattering study of the ferroelectric phase transition 
of Sr$_{0.61}$Ba$_{0.39}$Nb$_2$O$_6$. We show that despite strong and relevant chemical disorder this crystal largely behaves as an ordinary 
ferroelectric. 
 
Strontium Barium Niobate, Sr$_x$Ba$_{1-x}$Nb$_2$O$_6$ (SBN), is one of a group of disordered ferroelectrics with the tetragonal tungsten 
bronze structure~\cite{lines1977principles}. The dielectric anomaly associated with the phase transition in SBN is strongly affected by 
the relative amount of Sr/Ba ions. At high concentrations of Sr the dielectric anomaly in SBN occurs at lower temperatures, extends over two hundreds degrees 
and shows pronounced frequency dispersion. Upon increasing the Ba content the peak in the dielectric constant shifts towards higher temperature, becomes 
sharper and its dispersion nearly vanishes.  This suggests that the higher the Sr/Ba ratio the closer the shape 
of $\varepsilon '$ of SBN to that of relaxor ferroelectrics. 
Contrary to the observations for conventional cubic relaxors a spontaneous polarization $P_s$ develops in SBN as a result of the 
phase transition. 
This polarization is however unusual in several aspects. The value of $P_s$ at saturation and the critical 
exponent $\beta$ depend on how the poled state is prepared~\cite{PhysRevB_72_134105},\cite{voelker_114112}. 
Moreover, both rejuvenation and memory effects are observed in the ferroelectric state of 
SBN~\cite{PhysRevB_72_134105},\cite{PhaseTransitions_80_131}. Both these properties are characteristic of glasses rather than that of the ordered crystals. 
These observations for SBN have yet to be explained. 

\section{Crystal Structure and the Experimental Details}
\label{experiment}
The units cell of Sr$_x$Ba$_{1-x}$Nb$_2$O$_6$ is tetragonal for 0.25 $<$ x $<$ 0.75. Above T$_\mathrm{c}$ the space group of SBN 
is $4/mmm$, while below the phase transition temperature the space group is $4mm$ because the inversion centre is lost in the 
ferroelectric phase. The unit cell contains 5 formula units and the 5 Sr/Ba ions 
are distributed over 6 different sites. The structure is a network of distorted Nb/O octahedra connected together so that there are pentagonal, square and 
triangular tunnels which can be occupied by the Sr/Ba ions. These two ions have considerably different ionic radii~\cite{Shannon:a12967} and 
are distributed in the pentagonal tunnels, while only the Sr ions are randomly distributed in the square tunnels. As a result, this structure has some 
of the Sr ions and all of the Ba ions randomly distributed in the pentagonal tunnels and the remaining Sr ions are randomly located in the square tunnels. Thus SBN 
is disordered both chemically and in the position of the Sr/Ba ions. Since the Sr$^{2+}$ and Ba$^{2+}$ ions have formally the same charge there are no strong 
electrostatic forces related with the chemical randomness of the Sr and Ba ions. This is in contrast with the cubic relaxors, such as 
PbMg$_{1/3}$Nb$_{2/3}$O$_3$ (PMN) where the disordered ions Mg$^{2+}$/Nb$^{5+}$ have different charges.

We report below a neutron scattering study of Sr$_{0.61}$Ba$_{0.39}$Nb$_2$O$_6$(SBN-61) for which the transition temperature is T$_\mathrm{c}\sim$~350~K. 
The experiments were performed with the cold neutron 3-axis spectrometer TASP~\cite{Semadeni2001152}, 
situated at the end of a curved guide at the SINQ facility (PSI, Switzerland). The energy of the scattered neutrons was kept fixed to 8.04~meV and 
a PG filter was installed in front of the analyzer. Most of the data was collected with the collimation in the horizontal plane from reactor to detector 
as open-80$'$-$sample$-80$'$-80$'$, giving the energy resolution of 0.40 meV. Some data was taken with a tighter collimation: open-20$'$-$sample$-20$'$-20$'$, 
improving the resolution to 0.2 meV. A part of the data was collected with the 
wave-vector transfers near to the (001) Bragg reflection, while the majority of the data was obtained with the wave-vector transfers near the (002) 
Bragg peak. These Bragg positions were chosen because they have very different neutron structure factors ($|F|_{001}=9.6$ and $|F|_{002}=26.9$), 
while the spontaneous dielectric polarisation of SBN-61 is along the $<$0,0,1$>$ direction. 

The large single crystal of SBN-61 had dimensions of $2\times1.5\times1.5$ cm$^3$ and there was no sign of cracking. The mosaic spread was 
within the resolution of the spectrometer used for these experiments. The sample was mounted so that the $<$0,0,1$>$ and the $<$1,1,0$>$ directions 
defined the scattering plane. The lattice parameters of SBN-61 at T=300 K are $a = b= 12.488$ \AA, and $c = 3.938$ \AA. The sample was mounted in a displex 
refrigerator that enabled the temperature to be controlled between 20 K and 480 K. Higher temperature data was obtained by using a furnace that allowed for 
temperature to be as high as 800 K.

\section{Temperature dependence of the excitations} 
\label{energy}
%
%%%%%%%%%%%%%%%%%%%%%%%%%%%%%%%%%%%%%%%%%%%%%%%%%%%%%%%%%%%%%%%%%%%%%%%%%%%%%%%%%%%%%%%%%%%%%%%%%%%%%%%%%%%%%%
\begin{figure}
\centering
\includegraphics[width=0.5\columnwidth]{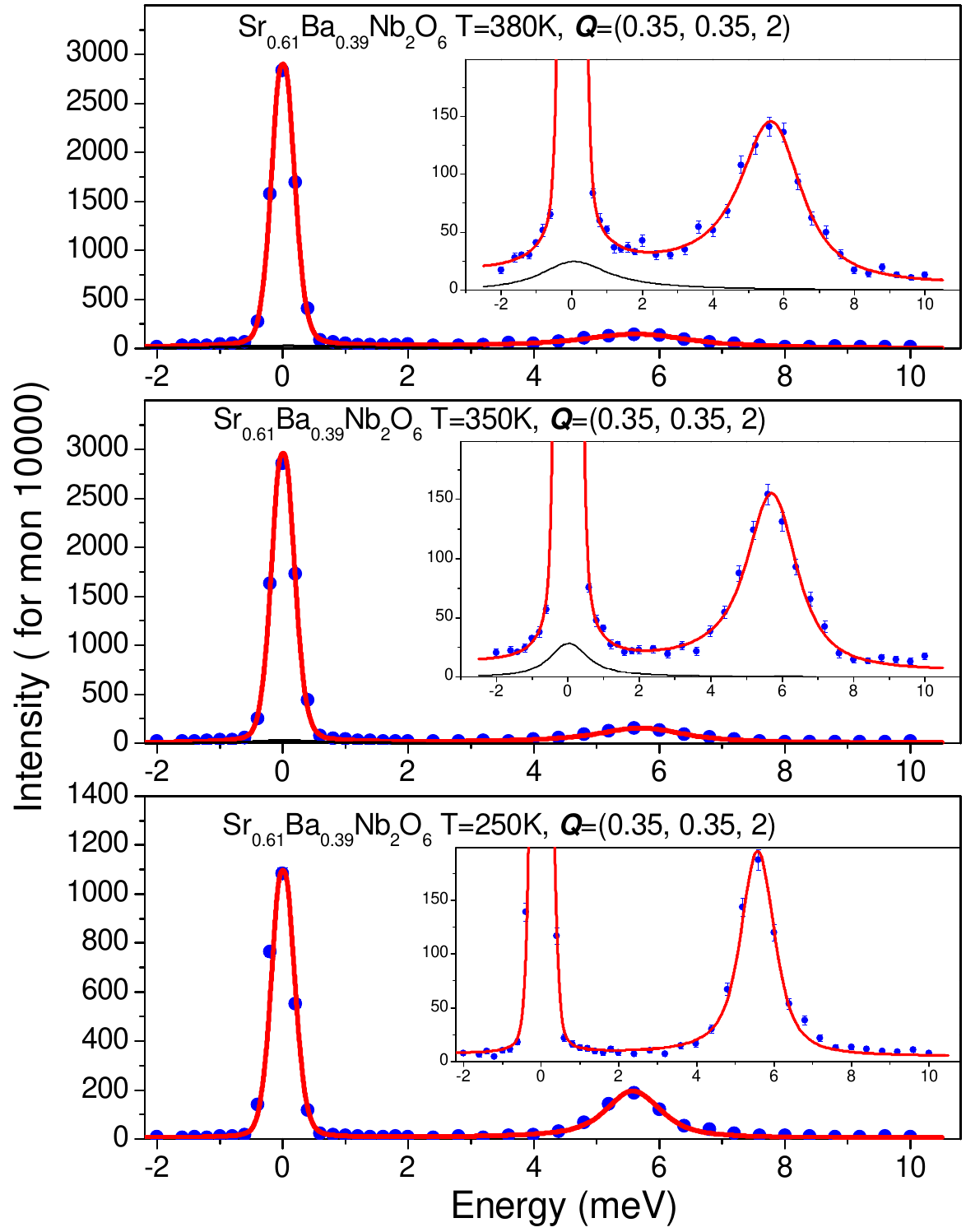}
\caption{Typical neutron spectra from SBN-61 taken in paraelectric and ferroelectric phases. Circles stand for the experimental data. 
Red line results from the fits as described in the text. Thinner solid line denotes weak dynamic component of the diffuse scattering.
         The insets emphasize the low-intensity part of the spectra.} 
  \label{figure01}
  \end{figure}
%%%%%%%%%%%%%%%%%%%%%%%%%%%%%%%%%%%%%%%%%%%%%%%%%%%%%%%%%%%%%%%%%%%

\noindent On approaching the ferroelectric phase transition the susceptibility of the transverse polarization fluctuations should diverge. 
Since in SBN the polarization appears along the $c$- axis, we studied the transverse modes propagating 
along the $<$q, q, 0$>$ direction. Fig.~\ref{figure01} shows the neutron spectra for a wave-vector transfer of (0.35,~0.35,~2) 
at temperatures above, near, and below the 
phase transition T$_\mathrm{c}\sim$350~K. The data show two peaks, one of which is elastic scattering while the other is an inelastic peak at an 
energy that increases as the wave-vector transfer is increased. 

To quantitatively analyze the spectra we establish a minimal required model. 
Attempts were made to fit the data with a Gaussian peak centred at zero energy and a damped harmonic oscillator (DHO) describing the inelastic 
scattering. 
This model failed to describe the experimental data satisfactorily, 
especially at low energy transfers where there was considerably more scattering than that given by the model. The scattering function was therefore 
extended by including a Lorentzian peak with a variable energy width and centred on zero energy transfer:

\begin{equation}
\label{sf_phonon}
S(\mathbf{Q},\omega) = A_{EDS}(\mathbf{Q})\delta(\omega)+\frac{1}{\pi}(n(\omega)+1)\cdot Im(B(\mathbf{Q})_{DHO} 
\cdot \mathlarger{\chi}_{DHO}+C_{L}(\mathbf{Q})\cdot\mathlarger{\chi}_{L}) 
\end{equation} 
\noindent In Eq.~\ref{sf_phonon} the first term describes the elastic diffuse scattering (EDS). The energy width of this scattering is 
limited by the experimental resolution. The second and third terms are the phonon scattering and the energy-resolved component of 
the diffuse scattering (IDS). Both of these terms have a finite energy width and so they contain the detailed  balance 
factor namely $(n(\omega)+1)=(1-exp[-\hbar\omega/(k_bT)])^{-1}$, $\mathlarger{\chi}_{DHO}=(\omega_q^2-i\gamma_q\omega-\omega^2)^{-1}$ 
and $\mathlarger{\chi}_{L}=(1-i\omega/\Gamma_L)^{-1}$ are the wavevector-dependent susceptibilities of the phonon 
and of IDS. The damping $\gamma_q$, $\Gamma_q$, and the scale factors 
$A_{EDS}$, $B(\mathbf{Q})_{DHO}$, and $C_{L}(\mathbf{Q})$ were allowed to have any constant value for each different wave vector. 
The scattering function Eq.~\ref{sf_phonon} was convoluted with the resolution function by using 
ResLib4.2 library~\cite{reslib} and a constant background was added. The best fits were obtained for a locally constant dispersion of the TA phonon. 
The fitted results gave a good description of the data as shown in Fig.~\ref{figure01}. 
%
%%%%%%%%%%%%%%%%%%%%%%%%%%%%%%%%%%%%%%%%%%%%%%%%%%%%%%%%%%%%%%%%%%%%%%%%%%%%%%%%%%%%%%%%%%%%%%%%%%%%%%%%%%%%%%
\begin{figure}
\centering
\includegraphics[width=0.5\columnwidth]{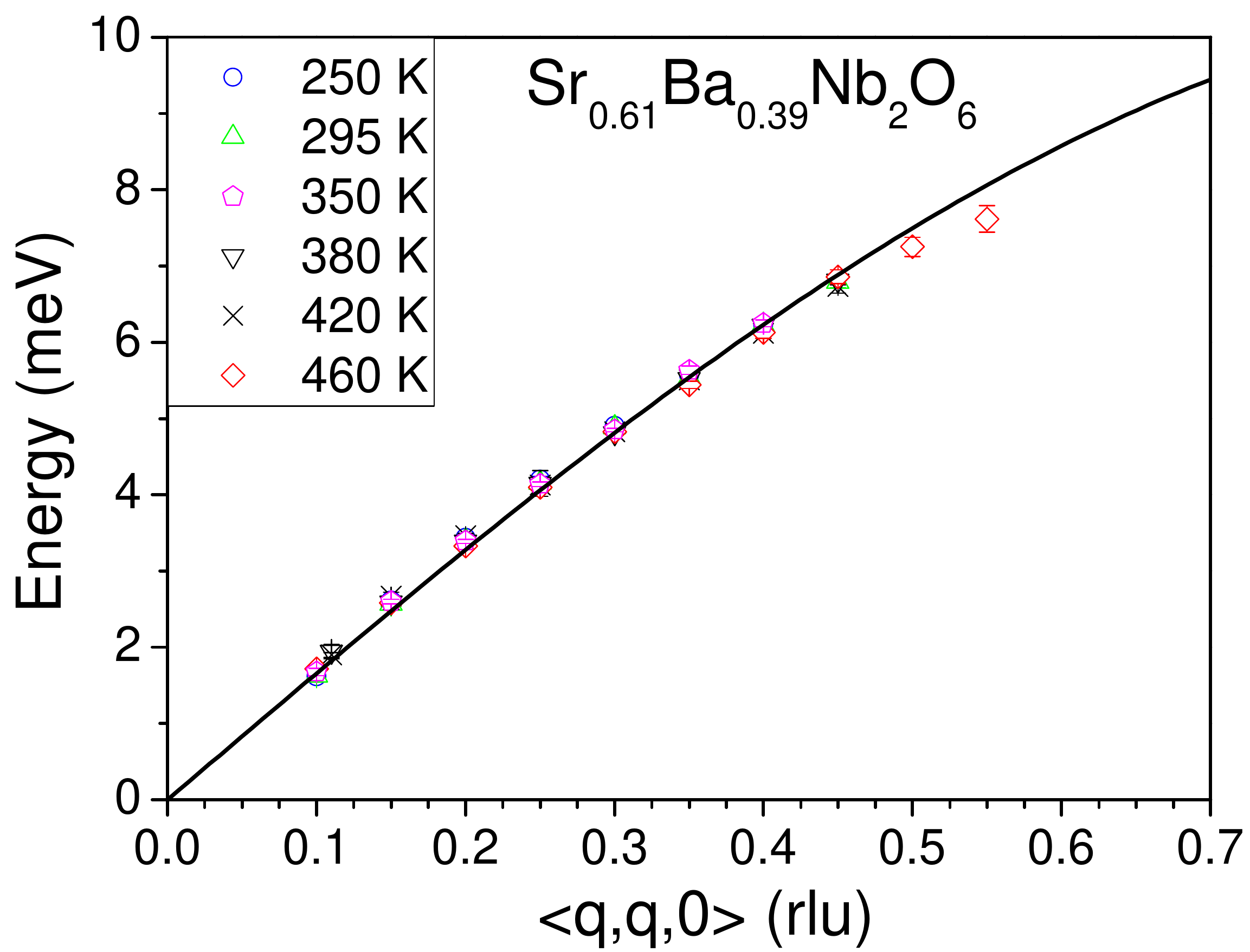}
\caption{Dispersion of the transverse acoustic phonon in SBN-61 propagating along the $<$1, 1, 0$>$ direction as a function of temperature.}
  \label{figure02}
  \end{figure}
%%%%%%%%%%%%%%%%%%%%%%%%%%%%%%%%%%%%%%%%%%%%%%%%%%%%%%%%%%%%%%%%%%%

We now qualitatively describe the behavior of each component of the neutron spectra for SBN-61. 
A weak IDS does not exhibit any regular dependence as a function of reduced wavevector q. 
As a function of temperature its intensity gradually decreases (see Fig.~\ref{figure01}). Such behavior suggests this scattering is not 
related to the ferroelectric transition of SBN-61 and we are not going to further consider the properties of this excitation.   
Due to the relatively low tetragonal symmetry the phonons in SBN are strongly anisotropic. Because of this a reliable parametrization of 
the data as a function of the wavevector is a very difficult problem both above and below T$_\mathrm{c}$. 
Fortunately, a good description of the phonon could be obtained by replacing a DHO with a Lorentzian form. Using this approach the 
dispersion of the TA phonon along the $<$q, q, 0$>$ direction was obtained and the result is shown in Fig.~\ref{figure02} for several temperatures. 
Within the experimental error, there is no change in the energy of the mode as the temperature is varied and it can be well approximated 
by $10.6\cdot \sin (0.5\pi\cdot q)$~meV. These measurements also did not reveal any scattering which could be associated with an soft optic phonon.

These results strongly suggest that the phonons do not play any role in the ferroelectric transition of SBN-61. 
In contrast, inspection of Fig.~\ref{figure01} shows that the elastic component of the diffuse scattering, EDS,  
significantly changes. For this reason we concentrate on the analysis of the EDS above (Section~\ref{dsabove}) 
and below (Section~\ref{dsbelow}) T$_\textrm{c}$ .

\section{Diffuse Scattering above T$_\textrm{c}$ and the Dielectric Anomaly}
\label{dsabove}
%
%%%%%%%%%%%%%%%%%%%%%%%%%%%%%%%%%%%%%%%%%%%%%%%%%%%%%%%%%%%%%%%%%%%%%%%%%%%%%%%%%%%%%%%%%%%%%%%%%%%%%%%%%%%%%%
\begin{figure}
\centering
\includegraphics[width=0.6\columnwidth]{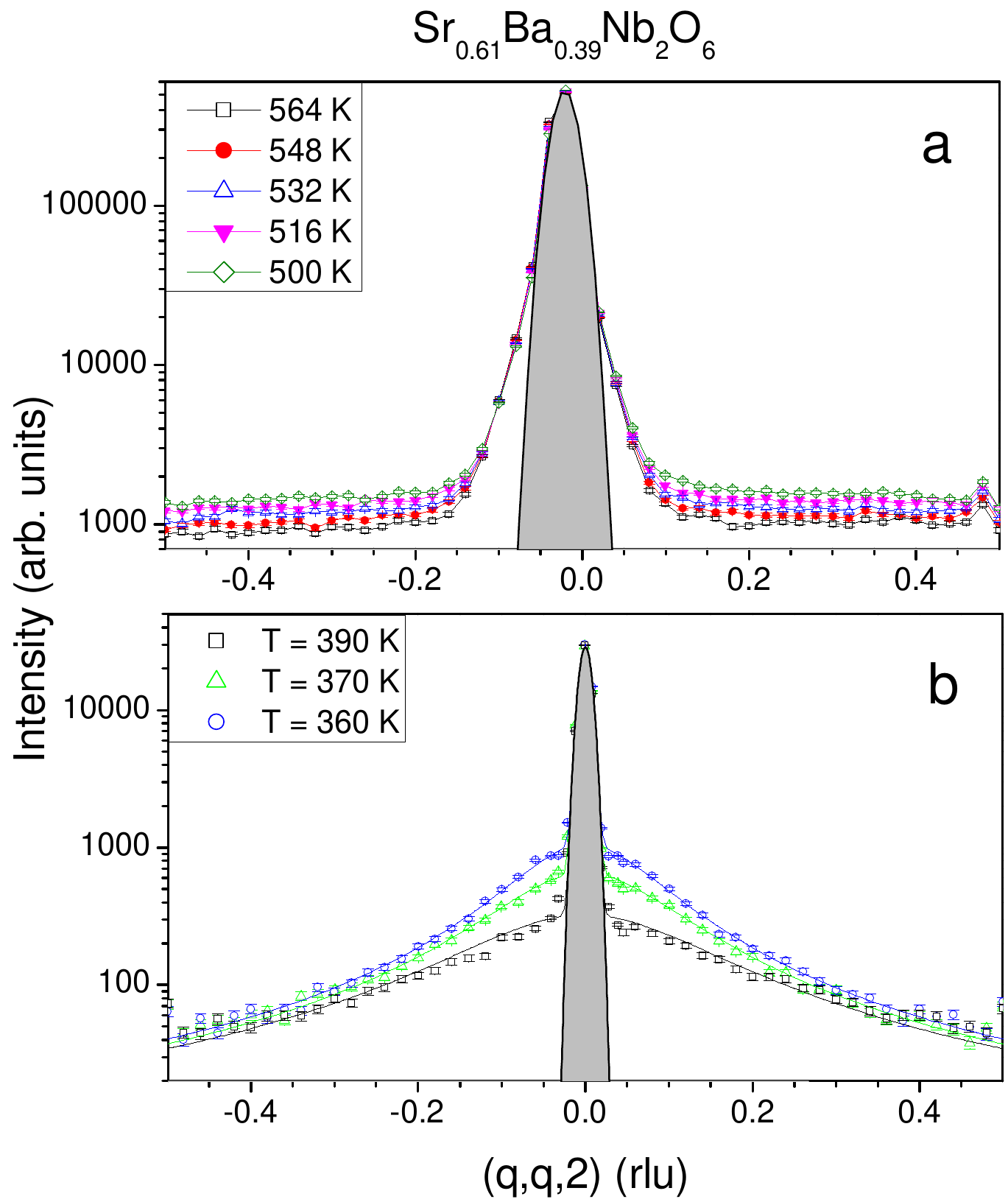}
\caption{Evolution of the DS in SBN-61 measured along the $<$1, 1, 0$>$ direction above T$_\textrm{c}$. The lines in 
Fig.~\ref{figure03}a are guide for the eye. The shaded areas in both panels of Fig.~\ref{figure03} emphasize the Bragg peaks and their widths  
the calculated with ResLib~4.2~\cite{reslib}. The data presented in Fig.~\ref{figure03}a is taken with collimation open-80$'$-80$'$-80$'$, 
whereas the high-resolution data presented in Fig.~\ref{figure03}b is obtained with open-20$'$-20$'$-20$'$.  Note the logarithmic scale. 
}
  \label{figure03}
  \end{figure}
%%%%%%%%%%%%%%%%%%%%%%%%%%%%%%%%%%%%%%%%%%%%%%%%%%%%%%%%%%%%%%%%%%%

\noindent 
The elastic neutron scattering was measured around 
the (0, 0, 1) and (0, 0, 2) Bragg reflections by performing scans along the $<$q, q, 0$>$, transverse, direction. 
At high temperatures, above about 450~K, the scattering consists of a sharp Bragg peak with a nearly flat background that is small and decreases only slightly 
as the temperature increases as shown in Fig.~\ref{figure03}a. Below $\sim$~450~K the elastic scattering contains a Bragg peak and a 
broader component that has a maximum at the $q=0$ position, as seen in Fig.~\ref{figure03}b. This additional contribution is clearly different from IDS as 
it rapidly increases in intensity upon cooling. Thus, this broader component of the elastic scattering is the EDS associated with the slow fluctuations 
and produces the elastic peak in the spectra shown in Fig.~\ref{figure01}.  

%
%%%%%%%%%%%%%%%%%%%%%%%%%%%%%%%%%%%%%%%%%%%%%%%%%%%%%%%%%%%%%%%%%%%%%%%%%%%%%%%%%%%%%%%%%%%%%%%%%%%%%%%%%%%%%%
\begin{figure}
\centering
\includegraphics[width=0.7\columnwidth]{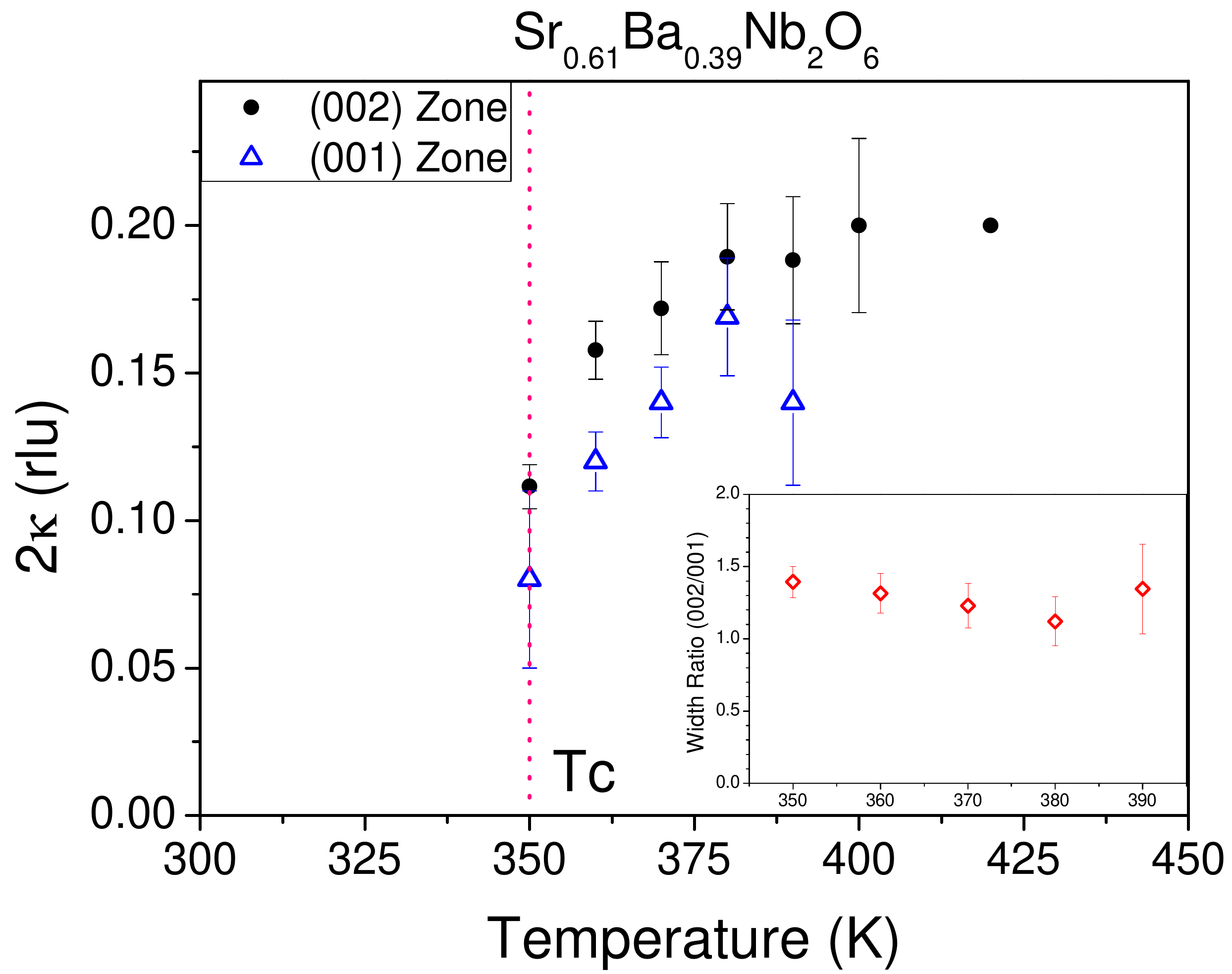}
\caption{The temperature dependence of inverse of the correlation length of the DS along the [1, 1, 0] direction. 
The inset show the ratio of $\kappa$ obtained from the scans near the (002) and the (001) Bragg peaks. Vertical red line denotes T$_\mathrm{c}$.

}
  \label{figure04}
  \end{figure}
%%%%%%%%%%%%%%%%%%%%%%%%%%%%%%%%%%%%%%%%%%%%%%%%%%%%%%%%%%%%%%%%%%%
A critical scattering above T$_\textrm{c}$ is commonly described by a Lorentzian function.  We shall assume that the fluctuations causing 
this EDS for SBN are dynamic, although they are very slow as their energy width has not been resolved. 
In this case the neutron intensity I$(\mathbf{Q},\omega)$ is proportional to the scattering function 
\begin{equation}
\label{sf_qe1}
S_{EDS}(\mathbf{Q},\omega)=\frac{(n(\omega)+1)}{\pi}\cdot Im(\mathlarger{\chi}_{EDS}(\mathbf{Q},\omega)) 
\end{equation} 
with susceptibility $\chi_{EDS}(\mathbf{Q},\omega)$ given by
\begin{equation}
\label{sf_qe2}
\chi_{EDS}(\mathbf{Q},\omega)=\frac{\mathlarger{\chi}(q=0,T)}{1-(\mathbf{Q}-\mathbf{\tau})^2\xi^2}(1-i\omega/\Gamma_q)^{-1}
\end{equation} 
\noindent In Eq.~\ref{sf_qe1}-\ref{sf_qe2} $\mathbf{Q}$ is the total wavevector transfer, $\mathbf{\tau}=\mathbf{Q} \pm \mathbf{q}$ is the appropriate 
reciprocal lattice point, $\chi(q=0,T)$ is the susceptibility at the $q=0$, $\xi=1/\kappa$ is the correlation length, and 
$\Gamma_q$ is the wavevector-dependent damping. This scattering is integrated over the energy transfer because the spectrometer resolution 
is much larger than the relevant energy scales of the EDS. The intensity is then adequately approximated by 
\begin{equation}
\label{sf_qe3}
I_{EDS}\sim\chi(q=0,T)\cdot T \cdot \frac{\kappa}{\kappa^2+q^2}
\end{equation} 
 
The scattering profiles shown in Fig.~\ref{figure03}b were fitted to a Gaussian Bragg peak together with an additional peak fitted to a Lorentzian curve 
convoluted with the Gaussian width that was determined from the Bragg reflections to be approximately the resolution width. The results for the 
width of the diffuse scattering are shown in Fig.~\ref{figure04}. They are difficult to extract reliably above a temperature of 400 K and below 350 K. A 
surprising feature of the results is that the width does not decrease to zero when the sample 
is cooled from a temperature of 400 K to T$_\textrm{c}$. At most phase transitions the 
width of the critical scattering decreases 
to a very small value. Another unexpected result is that the widths of the diffuse scattering observed near the (0, 0, 2) Bragg reflection is wider 
by approximately 30\%, than those observed near the (0, 0, 1) Bragg reflection (see inset to Fig.~\ref{figure04}). This difference is not related 
to possible gradients in the composition of the sample as 
the radial, transverse and longitudinal scans through the (1, 1, 0) and the (0, 0, 2) reflections appear to be of the resolution width and their 
shapes do not show appreciable 
changes with the temperature. The ratio of the widths about 1.3  and is unexpected because for most structural phase transitions it would be unity. On the other hand, 
if the increase of the width is caused by the strain, the ratio of the 
widths would be as high as  two. Our result is intermediate and suggests that somewhat different components of the ionic displacements 
contribute to the critical scattering around the (0, 0, 1) and the (0, 0, 2) Bragg peaks. 

\begin{figure}
\centering
\includegraphics[width=0.6\columnwidth]{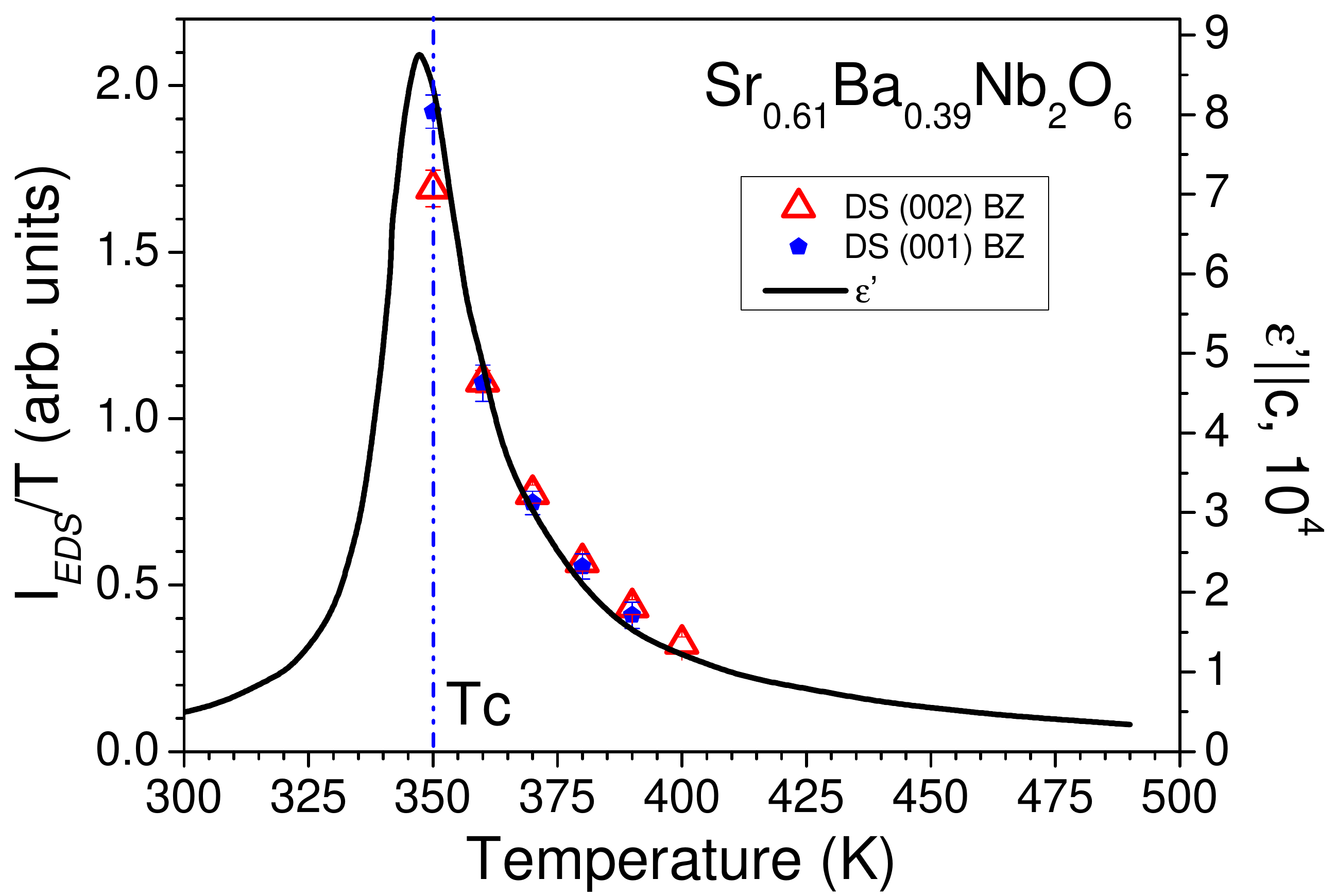}
\caption{The temperature dependence of the integrated susceptibility associated with the DS the (002) 
and the (001) Bragg peaks. The good agreement shows that the temperature dependence of the scattering has the same origin as the anomaly in the dielectric constant. 
The ratio of the intensities I(0,0,2)/I(0,0,1)$\sim7.4$ is equal to the ratio of the squared structure factors for the respective Bragg peaks. 
The dielectric data is taken from Ref.~\cite{Lukasiewicz20081464}. 
}
  \label{figure05}
  \end{figure}

One of the main results of this study is the similarity in the divergence of the EDS and dielectric anomaly observed in the paraelectric phase. 
The temperature dependence of the integrated intensity of the EDS scattering $I_{EDS}/T\sim\chi(q=0,T)$ is shown in Fig.~\ref{figure05} 
where it is compared with the real part of the low-frequency dielectric susceptibility measured along the ferroelectric $c$-axis, $\varepsilon' || c$. 
The agreement in the temperature dependences suggests that the diverging dielectric constant and the neutron elastic diffuse scattering have the same origin. 
Moreover, as no important changes were observed in the inelastic scattering probed near the (0, 0, 1) and the (0, 0, 2) Bragg peaks, it becomes apparent 
that slow fluctuations drive the phase transition of SBN-61 and are responsible for the ferroelectric anomaly. Based on the energy resolution in our experiments 
the upper time-scale of these fluctuations is $\sim10^{-11}$~s.

The inverse correlation length of the diffuse scattering obtained in our study is larger, on average, by a factor as high as 4 than that deduced 
for SBN-60~\cite{pss_55_334}. The major reason for this discrepancy is 
that in our experiments a much better signal-to-noise ratio was achieved. Indeed, Borisov et al.~\cite{pss_55_334} could follow the DS up to 
the wavevector transfer of q$\sim$0.18 rlu. Above this value the DS became indistinguishable from background~\cite{pss_55_334}. 
An incorrect determination of the background level creates fundamental problems in the model chosen to describe the scattering. 
Our parametrization of the neutron diffuse scattering in SBN is more reliable. 
%
%%%%%%%%%%%%%%%%%%%%%%%%%%%%%%%%%%%%%%%%%%%%%%%%%%%%%%%%%%%%%%%%%%%%%%%%%%%%%%%%%%%%%%%%%%%%%%%%%%%%%%%%%%%%%%
%%%%%%%%%%%%%%%%%%%%%%%%%%%%%%%%%%%%%%%%%%%%%%%%%%%%%%%%%%%%%%%%%%%
%
%
\section{Diffuse Scattering from Ferroelectric Domain Walls}
\label{dsbelow}

The elastic scattering in the ferroelectric phase of SBN is more difficult to analyse.  
Below the phase transition temperature, it becomes complicated to distinguish the Bragg peaks and the diffuse scattering since the latter becomes more 
intense close to $q=0$. Furthermore, one should expect at least two contributions to the diffuse scattering below, but near T$_\textrm{c}$. 
The first contribution comes from the critical scattering. As we have shown for T$>$T$_\textrm{c}$, 
the intensity of the DS follows well $\varepsilon '$ and a similar behaviour is expected for T$<$T$_\textrm{c}$. An inspection of Fig.~\ref{figure05} 
suggests that the critical scattering should become negligible at T$\le$300~K. Second,  
below T$_\textrm{c}$ the sample breaks up into ferroelectric domains, which causes intense diffuse scattering from the domain walls. Finally, 
due to the relief of extinction, a change in the Bragg intensity near T$_\textrm{c}$ is expected. In order to reliably describe the data, the temperature 
range was restricted to T$\le300$~K. In this regime we can expect the EDS to be dominated by the scattering from the the ferroelectric domain walls. 

The distribution of the EDS in the [H, H, L] plane was measured by scanning along several directions and around a number of the Bragg peaks. 
In particular, longitudinal scans were performed around the (0, 0, 1), (0, 0, 2), and (2, 2, 0) Bragg positions. 
The EDS is not observed in this geometry. Also, the scans along the $<$q, q, 0$>$, $<$0, 0, q$>$, and $<$q, q, q$>$ were performed in the vicinity of the (1, 1, 1), 
and (1, 1, 2) Bragg peaks. The diffuse scattering was detected only for the $<$q, q, 0$>$ directions. Our observations are in agreement with previous reports on 
the distribution of the DS in SBN~\cite{PSSB:PSSB2221130124},~\cite{PSSA:PSSA2211030114},~\cite{pss_55_334} and show the diffuse scattering is sharp 
along the $<$0,0,q$>$ direction.

Here we recall the results for the scattering from domain walls~\cite{PhysRevLett36806},\cite{bruce_kdp},\cite{andrews_kdp},\cite{PhysRevB193630} and 
introduce the necessary modifications. If no electric field is applied to the sample when the temperature passes through T$_\textrm{c}$ from above, 
the structure breaks up into ferroelectric domains. 
For a uniaxial ferroelectric a plausible assumption is that the adjacent domains with opposite polarizations along the $c$-axis, thus representing the 
so-called 180$^{\circ}$ structure. In this way the problem is reduced to the scattering by the walls with the normals perpendicular to the $c$-axis. In order to obtain 
an analytical expression for the scattering function, a further simplification is helpful. We shall assume that the spontaneous polarization along an arbitrary 
direction $x$, $P_x$, is proportional to the displacement field  $P_x \sim u(x)$ with $u(x)=u_0\tanh(x/\lambda)$. This leads to the 
scattering function having the form 

\begin{equation}
\label{dw1}
S(q_x)=\frac{D\lambda^2}{\sinh^2(q_x\lambda/2)} 
\end{equation} 
\noindent In this approach, $\lambda$ is a half of the average thickness of the domain walls along $x$ direction, and $D$ is a constant proportional to the density of the 
domain walls. 
%
%%%%%%%%%%%%%%%%%%%%%%%%%%%%%%%%%%%%%%%%%%%%%%%%%%%%%%%%%%%%%%%%%%%%%%%%%%%%%%%%%%%%%%%%%%%%%%%%%%%%%%%%%%%%%%
\begin{figure}
\centering
\includegraphics[width=0.6\columnwidth]{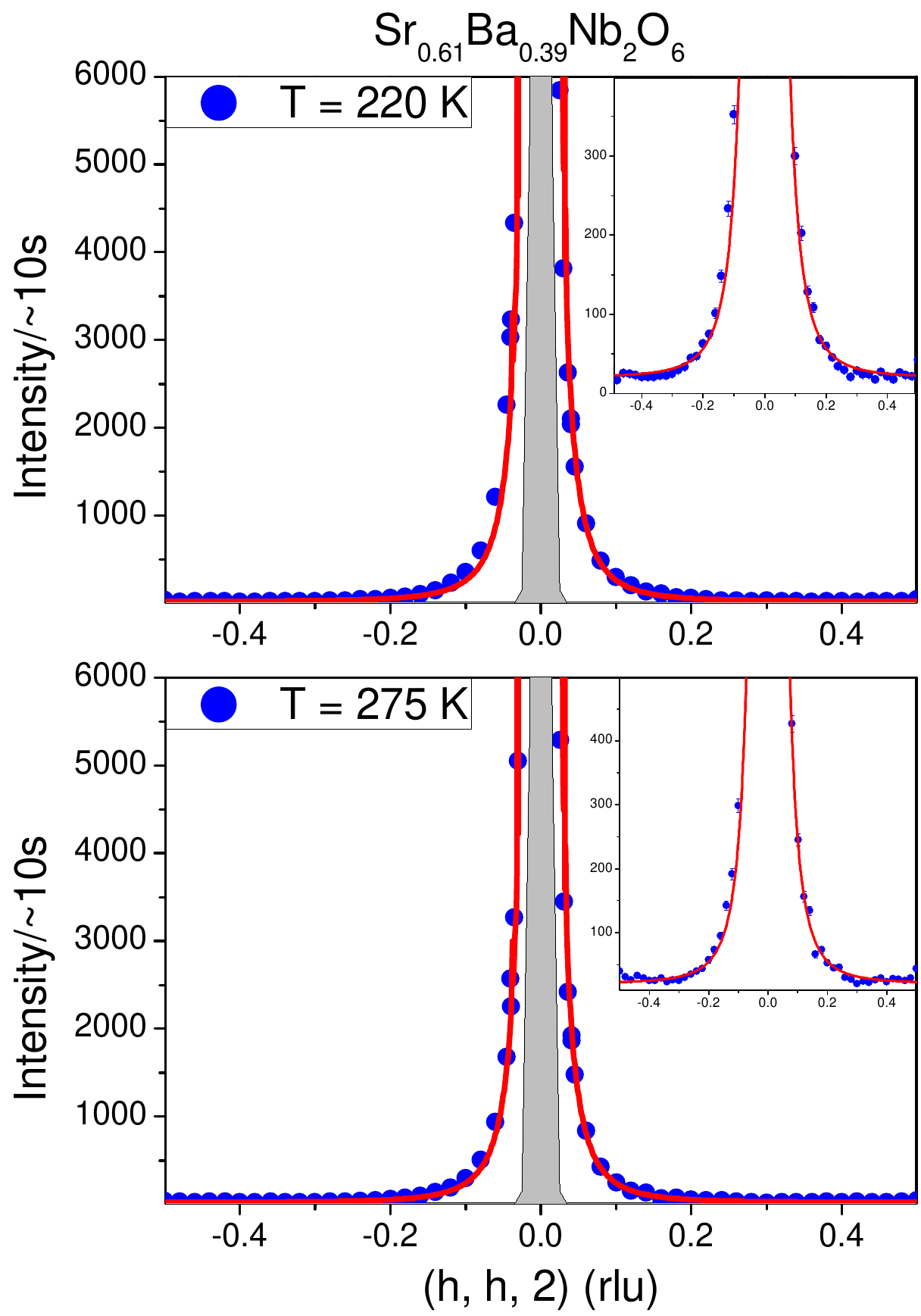}
\caption{The EDS below the phase transition in SBN-61 with fits to the domain wall model. The data are shown by blue circles and 
the fits to the Eq.\ref{dw3} are given by red lines. The insets emphasize the low-intensity parts of the scans. 
The shaded areas denote the portion of the scans where the intensity would be influenced by the Bragg peak.
The data was taken with collimation as open-20$'$-20$'$-20$'$.  
}
  \label{figure06}
  \end{figure}
%%%%%%%%%%%%%%%%%%%%%%%%%%%%%%%%%%%%%%%%%%%%%%%%%%%%%%%%%%%%%%%%%%%

In order to properly account for the effects of instrument resolution, the model represented by Eq. (\ref{dw1}) has to be extended 
to 4D $(\mathbf{Q},\omega)$ space. To this 
end Eq. (\ref{dw1}) can be rewritten in the following way: 

\begin{equation}
\label{dw2}
S(\mathbf{Q},\omega)=\frac{D\lambda^2}{\sinh^2(q_x\lambda/2)}\cdot\delta(q_y)\cdot\delta(q_z)\cdot\delta(\omega)  
\end{equation} 

\begin{figure}
\centering
\includegraphics[width=0.6\columnwidth]{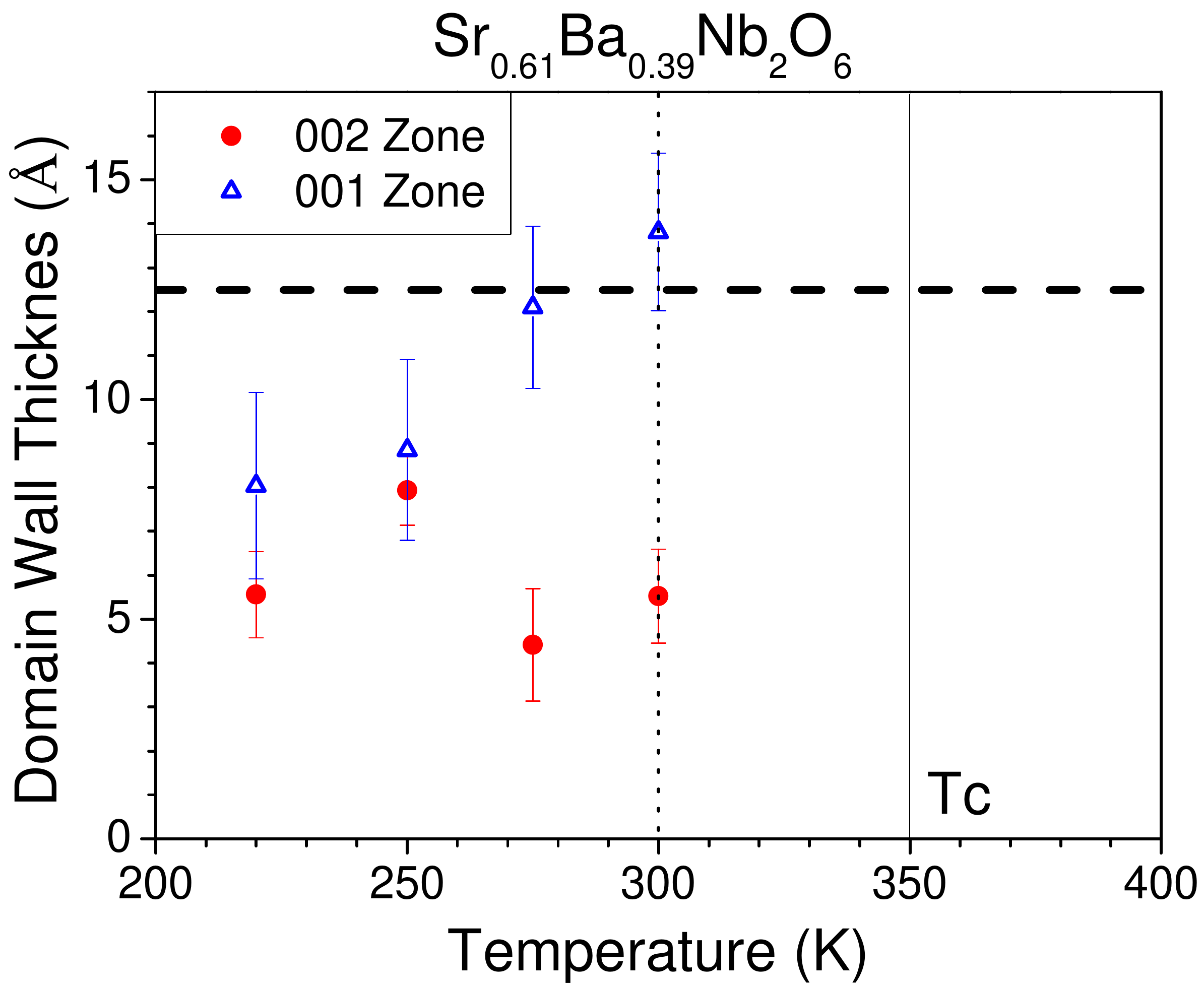}
\caption{The width of the ferroelectric domain walls in SBN-61 in the temperature range 220~K-300~K. The vertical lines denote the 
phase transition T$_\textrm{c}$ and the temperature where the critical contribution to the DS becomes negligible. The 
horizontal line shows the value of the lattice constant. 
}
  \label{figure07}
  \end{figure}

\noindent The $\delta$- functions in Eq.~(\ref{dw2}) are now considered. The $\delta(\omega)$ is directly applicable for the convolution as the EDS 
is resolution-limited in energy. The EDS along the $<$0, 0, q$>$ direction is sharp and appears to be also resolution limited. For this reason the $\delta(q_z)$ 
in Eq.~(\ref{dw2}) is approximated by a sharp (much narrower than the calculated resolution) Lorentzian.  Finally, we assume that the domain walls are 
randomly distributed in the $a-b$ plane of the crystal and hence do not produce interference effects. This allows us to average the 
$1/{\sinh^2(0.5q_x\lambda)}\cdot\delta(q_y)$ term in Eq.~(\ref{dw2}) leading to the following scattering function:
\begin{equation}
\label{dw3}
S(\mathbf{Q},\omega)=D\frac{1}{|q|}\frac{\lambda^2}{\sinh^2(q\lambda/2)}\frac{\kappa_z}{q^2+\kappa_z^2}\cdot\delta(\omega)  
\end{equation} 
\noindent In Eq.~(\ref{dw3}) the reduced wavevector $q$ runs from respective Bragg peak, $\mathbf{q}=\mathbf{Q} \pm \mathbf{\tau}$, 
and in our treatment $\kappa_z$ was fixed to the value 0.0005 rlu. Eq.~\ref{dw3} was convoluted with the resolution function 
of the spectrometer and the best fit results are shown in Fig.~\ref{figure06}.  The points in the scans 
that are clearly a part of the Bragg peak (shaded areas in Figs.~\ref{figure03}b and~\ref{figure06}) were excluded from the fitted range. 
The model reproduces the data taken around the Bragg peaks (0,0,1)  
and (0,0,2) in the temperature range 220~K -- 300~K. The values of 2$\lambda$ are shown 
in Fig.~\ref{figure07} and they do not exhibit significant temperature dependence. 
The wall thickness inferred for SBN-61 is close to the value of the lattice parameters. This result is in agreement with the generally expected 
(atomically) sharp domain walls in ferroelectrics and is similar to the results obtained for SBN-70 by Prokert~\cite{PSSB:PSSB2220870121} and for 
other uniaxial ferroelectrics~\cite{PhysRevLett36806},~\cite{PhysRevB193645},~\cite{philmaga45911}. 
%
%%%%%%%%%%%%%%%%%%%%%%%%%%%%%%%%%%%%%%%%%%%%%%%%%%%%%%%%%%%%%%%%%%%%%%%%%%%%%%%%%%%%%%%%%%%%%%%%%%%%%%%%%%%%%%
%%%%%%%%%%%%%%%%%%%%%%%%%%%%%%%%%%%%%%%%%%%%%%%%%%%%%%%%%%%%%%%%%%%
\section{Conclusions}
\label{concl}

\noindent 
We have studied the ferroelectric phase transition of SBN-61 by neutron scattering. Measurements of the transverse acoustic mode as a function of 
temperature showed very little change in the dispersion and did not give any evidence for the soft optic phonon.
We did observe critical scattering with a temperature dependence similar to that of the low-frequency dielectric constant. 
The time scale of this scattering is slower than $10^{-11}$~s. The wavevector width of the scattering reduces as the sample is cooled but does not 
reach zero at the ferroelectric phase transition. In the ferroelectric phase the neutron diffuse scattering could be  
described by scattering from sharp domain walls. Our results suggest that despite apparent chemical disorder SBN-61 behaves as ordinary order-disorder uniaxial 
ferroelectric with slow critical fluctuations. 

\section{Acknowledgments}
The experiments were performed at the SINQ facility at PSI and we wish to thank all those who assisted us. 
Excellent technical support from Markus Zolliker and Walter Latscha is greatly acknowledged. This work has been partly financially supported by the 
Ministry of Education and Science of Russian Federation, state contract no.16.513.12.3019.
and by the Swiss National Foundation for the Scientific Research within the NCCR MaNEP pool. 
\end{document}